# Towards a better understanding of the Anthropocene

*Ron W. Nielsen*

Retired nuclear scientist from Department of Nuclear Physics, The Australian National University, Canberra, ACT, 2601, Australia. E-mail: ronwnielsen@gmail.com, Tel.: +61-407-201-175.

*(1) Results of analysis of new sets of anthropogenic data are presented. They confirm earlier results of similar analyses. The expected and inevitable massive deceleration of human-induced global change process is demonstrated as an ongoing phenomenon. Human activities and impacts on the Earth System, while in general remaining strong, are now systematically slowing down. (2) It is now clearly stated, by supporters of the concept of the Great Acceleration, that the so-called Great Acceleration is not acceleration and that this term is open to misinterpretation. (3) There was no systematic sharp increase in anthropogenic growth trajectories around 1950 CE (Common Era) or around any other recent time. (4) Close inspection of data suggests that the Anthropocene is not a geological epoch, but a historical event, also reflected as a geological event. (5) The Anthropocene has no convincingly determinable beginning. What is described as the Anthropocene appears to be an integral part of a long process transcending the Pleistocene and Holocene epochs, a natural continuation of the past developments of human interaction with nature. (6) Using the concept of the Jinji unconformity, the embryonic beginning of human (hominid) interaction with the environment could be placed at the time of the first use of stone tools millions of years ago. Gradually the intensity of this interaction was increasing and culminated monotonically in an event described as the Anthropocene.*

## Introduction

The Anthropocene is understood as the latest interval of time characterized by exceptionally strong human interaction with the environment, so strong that it seemed to suggest transition to a new geological epoch (Crutzen and Stoermer, 2000). A direct way to study the concept of the Anthropocene is by a close inspection of anthropogenic growth trajectories describing human activities and impacts. This kind of study has now been carried out (Nielsen, 2021, 2022) and it revealed that human activities and impacts are now systematically slowing down. This analysis also demonstrated that there was no systematic intensification of growth around 1950 CE (Common Era) or around any other recent time and suggested a new interpretation of the Anthropocene. The aim of the presented here document is to show results of a new analysis of anthropogenic data as compiled by Our World in Data (2023). Analysis of the total global anthropogenic sediment flux (Cooper et al., 2018) is also included.

## Method of analysis

The fundamental variable describing the intensity of growth is the growth rate defined as a gradient (for instance, annual increase) divided by the size of a growing entity, and it is usually expressed in percent per year. It is a generally well-understood quantity. It is a relative increase, i.e., increase over a certain time related to the size of a growing entity. The familiar examples of the growth rate are interest rate and annual percentage change. Growth rate is used to describe growth in savings, superannuation, debts, economic growth, the spread of infectious diseases and many other processes. If the growth rate is high, growth is fast. If the growth rate is small, growth is slow. If growth rate is zero, there is no growth. If growth rate is positive, growth is increasing. If growth rate is negative, growth is decreasing. Growth rate represents the driving force of growth.

Identification of decelerating trajectories is easy. If growth rate is systematically (on average) decreasing (i.e., if the driving force of growth is gradually decreasing), growth is decelerating (slowing down). If the aim of the inspection of data is to check whether growth is slowing down (decelerating), then analysis of the growth rate over a sufficiently long time is perfectly suitable for this purpose. If, in addition, the aim is to describe growth trajectories mathematically and to project growth, then the required information can be obtained by solving the following simple mathematical equation:

$$\frac{1}{S}\frac{dS}{dt} = F \ . \tag{1}$$

The left-hand side of this equation is the analytical definition of the growth rate, with $S$ being the size of the growing entity. The right-hand side is a mathematical description of a general trend of a given empirical growth rate, which can be expressed as a function of time or as a function of the size of the growing entity.

If empirical growth rate can be, on average, described by a straight line dependent on time, i.e., if

$$F = a + bt , \tag{2}$$

then the solution to the eqn (1) is

$$S = C\exp(at + 0.5bt^2) , \tag{3}$$

which can be also expressed as

$$S = \exp(a_0 + a_1 t + a_2 t^2) \ . \tag{4}$$

For a reliable projection of growth, mathematical description of growth rate should not depend strongly on the independent variable and the simplest representation is a straight line. If empirical growth rate (driving force) decreases on average linearly with time, then this type of force generates second-order exponential growth reaching a maximum at a certain fixed time. If empirical growth rate (driving force) decreases linearly with the size of a growing entity, then this type of driving force generates logistic growth leveling off at a certain maximum value. If empirical growth rate (driving force) decreases, on average, exponentially with time, then the corresponding growth trajectory is pseudo-logistic, also levelling off at a certain maximum value. For more information, see earlier publications (Nielsen, 2017a, 2021 Suppl.).

## New evidence of decelerations

### *Deceleration of the air transport*

Data presented in Fig. A1 (in the Appendix) show that, on average, growth rate for air transport was steadily decreasing, which means that the driving force of growth was steadily decreasing. The growth of the air transport was gradually slowing down. Growth trajectory generated by this linearly decreasing growth rate (driving force) is shown in Fig. A2. It is a solution of the eqn (1) for a straight line shown in Fig. A1. In this semilogarithmic display, growth trajectory is bending downwards showing again that the growth was gradually slowing down. If this trend is going to continue, it will reach a predicted maximum of 281,983 Mt-km (Megatons-km) in around 2036 CE.

### *Deceleration of $CO_2$ emissions from cement*

Data presented in Fig. A3 show the growth rate for the global annual emissions of carbon dioxide from the production of cement. This growth can be divided into two regions, before and after around 1950 CE. Empirical growth rate between 1900 CE and 1950 CE was strongly fluctuating. A possible explanation of these large fluctuations is large inaccuracies in the reported data. From around 1950 CE fluctuations are much smaller. On average, growth rate was decreasing in both regions. Growth was gradually decelerating all the time. There was no acceleration (intensification of growth) around 1950 CE.

Results presented in Fig. A4 show two mathematical trajectories compared with data. The overall trajectory is clearly bending downwards and thus describing a gradually decelerating growth over the whole time. It might be interesting to notice that the general trend of growth trajectory is not dictated by fluctuations and oscillations in the growth rate. The general trend of growth is dictated by the long-term trend of the growth rate.

Trajectory from around 1950 CE gives a more reliable projection of growth because it is based on the analysis of more accurate data. The predicted maximum is 2045 Mt of $CO_2$ in around 2045 CE.

### *Deceleration of the global emissions of $CO_2$*

Data presented in Fig. A5 show a gradual decrease of the growth rate of the global annual emissions of $CO_2$. Global emissions of $CO_2$ were slowing down. The corresponding growth trajectory generated by the linear representation of the growth rate shown in Fig. A5 is displayed in Fig. A6. The projected maximum is 37.53 Gt (giga tonnes) per year in 2025 CE.

### *Deceleration of the global consumption of electricity generated by fossil fuels*

Data presented in Fig. A7 show a gradual decrease of the growth rate of the global annual consumption of electricity generated by fossil fuels: coal, gas and oil. Global annual consumption of electricity generated by fossil fuels has been slowing down. Empirical growth trajectory is compared in Fig. A8 with the trajectory generated by the linear representation of the empirical growth rate shown in Fig. A7. If this trend is going to continue, global generation of electricity using fossil fuels is expected to reach a maximum of 19,853 TWh (terawatt hour) around 2045 CE.

### *Deceleration of the global consumption of energy generated by fossil fuels*

Growth rate for the global annual consumption of energy generated by fossil fuels is shown in Fig. A9. Consumption of energy generated by these sources has been gradually slowing down. Growth trajectory generated by a straight line shown in Fig. A9 is displayed in Fig. A10. If this trend is going to continue, consumption of energy generated by fossil fuels can be expected to reach a maximum of 150,529 TWh per year around 2040 CE.

### *Deceleration of the global annual production of meat*

Growth rate for the global annual production of meat is shown in Fig. A11. Growth rate has been steadily decreasing demonstrating that global annual production of meat has been steadily decelerating (slowing down). The corresponding growth trajectory generated by the straight line shown in Fig. A11 is shown in Fig. A12. Global annual production of meat is projected to reach a maximum of around 554 Mt per year around 2070 CE.

### *Deceleration of the global annual emissions of $SO_2$*

Data presented in Fig. A13 show empirical growth rate for the global annual emissions of $SO_2$. Growth rate was gradually decreasing. Annual emissions were slowing down. Growth trajectory for the global annual emissions of $SO_2$ is shown in Fig. A14. The calculated trajectory was generated by the straight line shown in Fig. A13. Annual emissions of this gas have already reached a maximum of 151.5 Mt/y in 1980 CE.

### *Deceleration of the anthropogenic sediment flux*

Data for the total global anthropogenic sediment flux include the following human activities: world coal production including overburden/waste; world mineral and metal production (excluding oil and gas) but including overburden/waste; world aggregates and cement production; world civil engineering earthworks; and world dredging. Growth rate for the total global flux is shown in Fig. A15. Growth rate was on average increasing before 1950 CE.

Growth was on average accelerating. From around 1950 CE, or soon after that year, growth rate was on average systematically decreasing. Growth trajectory was slowing down. Growth trajectories generated by the linear representations of the growth rate shown in Fig. 15 are shown in Fig. 16. Transition from the accelerating to the decelerating trajectory occurred soon after 1950 CE. The new decelerating trajectory is projected to reach a maximum of around 1,719,000 Mt/y around 2115 CE. However, it is possible that the maximum might be reached earlier because growth rate was decreasing faster than the average trend before around 2000 CE. The brief boosting of the growth rate that commenced just before 2000 CE appears to be followed by a fast decline. Extended data are needed for a better prediction of growth.

Predicted maxima for the discussed here trajectories are listed in Table 1. Nearly all of them are predicted to reach their respective maxima during the current century, most of them before 2050 CE.

*Table 1. Predicted maxima for the discussed growth trajectories*

| Growth trajectory | Maximum |
| --- | --- |
| Air transport | 2036 |
| $CO_2$ emissions from cement | 2045/2075 |
| Total global $CO_2$ emissions | 2025 |
| Electricity from fossil fuels | 2045 |
| Energy from fossil fuels | 2040 |
| Meat production | 2070 |
| Anthropogenic sediment flux | 2115 |

## Discussion

### *Deceleration*

Massive deceleration (slowdown) of anthropogenic activities and impacts, demonstrated by the mathematical analysis discussed here and in earlier publications (Nielsen, 2021, 2022), is in close agreement with discussions of Steffen at al. (2004), Hibbard et al. (2005) and Head et al. (2022). Conceptual diagram of the *ongoing* deceleration of global change process was first presented by Steffen et al. (2004) in their Fig. 3.69, reproduced here as Fig. 1 and by Head et al. (2022) in their Fig. 3. The original diagram of Steffen et al. (2004) was shown it in conjunction with the now well-known series of diagrams presented by them in their Figs 3.66 and 3.67, which in their updated form were published by Broadgate et al. (2014), the diagrams which inspired the concept of the so-called "Great Acceleration". These diagrams illustrate the recently experienced rapid increase and even initial accelerations of human activities and impacts.

Steffen et al. (2004) reminded that fast-increasing trajectories cannot increase indefinitely and that at a certain time they have to be decelerated (slowed down). Decelerating trajectories can then follow any of the following three pathways. They can (A) level off at a certain maximum value, (B) reach a localised maximum and start to decrease gradually in a controllable or tolerable way, or (C) be abruptly terminated. It can be shown that sustainable trajectories (A) and (B) are prompted by driving forces decreasing to zero. For the trajectory (A), the driving force decreases asymptotically to zero, or equivalently, in its simplest form, linearly with the size of the growing entity. It is a logistic type of growth. For the trajectory (B), it decreases to zero at a certain fixed time, and in its simplest form can be described as the second-order exponential growth. All three of them are prompted by forces, which are becoming weaker, either gradually, for the trajectories (A) and (B), or abruptly, for the trajectory (C).

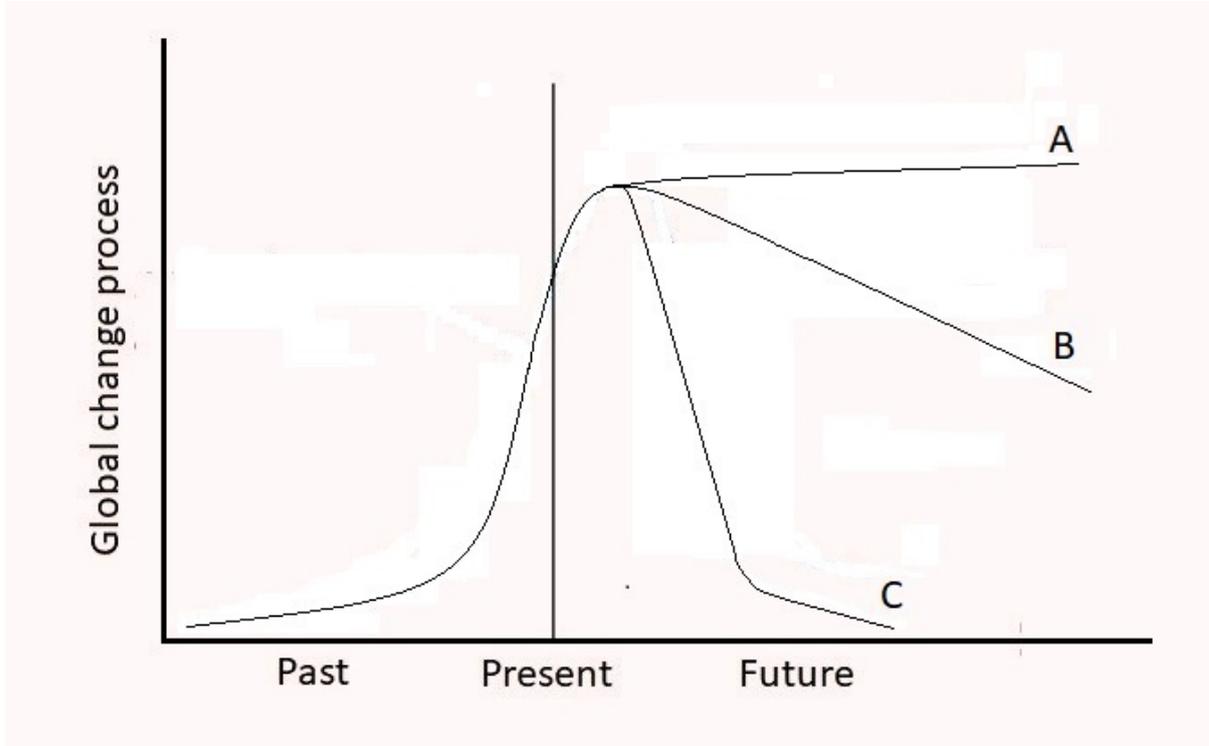

*Figure 1. Conceptual diagram of a massive deceleration of global change process based on Fig. 3.69 published by Steffen et al. (2004). This anticipated systematic deceleration has now been confirmed here and in earlier publications (Nielsen, 2021, 2022).*

The fundamental tenet of the concept of the Anthropocene is that global change process is caused now predominantly by humans. Consequently, the diagram presented in Fig. 1, showing deceleration (slowdown) of global change process, describes the deceleration of human activities and impacts. It is clearly a *massive* deceleration because global change process is obviously caused by a wide range of human activities, all described by a wide range of anthropogenic growth trajectories already analysed and yet to be analysed or at least identified.

What Steffen et al. (2004) presented in their Fig. 3.69 (reproduced here as Fig. 1) was not a prediction based on a study of the mechanism of growth, but only their conceptual interpretation of reality. They were the first to point out that deceleration of anthropogenic trajectories was inevitable. They were also the first to illustrate this process of deceleration by a diagram reproduced here as Fig. 1. Furthermore, they were also the first to publish their iconic diagrams (see Steffen et al., 2004, their Figs 3.66 and 3.67) that played a vital role in the demonstration of the anticipated massive deceleration, quickly corroborated by the analysis of data listed by Syvitski et al. (2020) and now by new data discussed in this document.

As Steffen et al. (2004) presented in their diagram, systematic and massive deceleration of anthropogenic activities and impacts responsible for global change process is now in progress. Its beginning is now well in the past. According to their diagram, the expected beginning of the massive deceleration was already in the past in 2004 CE. Analysis of data demonstrated that many anthropogenic trajectories were decelerating over the entire range of data, commencing well before 1950 CE and certainly well before 2004 CE, and that they continue to decelerate. However, many of them started to decelerate soon after 1950 CE or generally in the second half of the 20th century. It is a massive deceleration because nearly all anthropogenic growth trajectories, so far analysed, are now decelerating. Analysis of anthropogenic trajectories shows how closely the hypothetical interpretation of Steffen et al. (2004) is confirmed by data.

Examples of decelerations of anthropogenic activities and impacts were discussed by Hibbard et al. (2005). They saw no conflict between the proposed by them concept of the so-called "Great Acceleration" and the observed by them examples of deceleration, because the so-called "Great Acceleration" is not acceleration and was never intended to be interpreted as acceleration (Steffen, 2017, 2018, private communication; Head et al., 2022). Hibbard et al. (2005) even suggested that the observed decelerations are an *emergent property* of the so-called "Great Acceleration". They are a part of the same process. This ongoing process could be perhaps better described as the Great Rapid Increase because it is the rapid increase in human activities, not just the initial acceleration, in some cases, that has been so damaging to the Earth System in the past 200 years, or so. Growth of human (hominid) population was accelerating for millions of years (Nielsen, 2017b) and was not causing any serious problem until its steady acceleration, combined with the steady acceleration in human activities, reached gradually a stage of a rapid increase, which is still continuing even though, in general, anthropogenic trajectories are now undergoing a massive deceleration. Systematic study of decelerations is essential for the correct understanding of the Anthropocene. It would be also interesting to find and to investigate possible examples of acceleration. Two such examples were identified and discussed in the first publication, which demonstrated the massive deceleration (Nielsen, 2021).

## *New interpretation of the Anthropocene*

Analysis of the growth of hominid population (Nielsen, 2017b) over the past 2,000,000 years demonstrated a generally gradual increase (see Fig. 2). Temporary and regional leaps and delays appear to have been averaging out and blending into a largely steady growth of human (hominid) population. Considering this feature and the expectation that human activities and impacts on the environment are, on average, strongly correlated with the size of human population (Waters et al., 2016), the trajectory presented in Fig. 2 indicates that human interaction with the environment was developing gradually and steadily over a long time. There is no evidence in these data that the Anthropocene emerged suddenly at any time.

Growth of human population including the associated growth of human activities and impacts on the environment was in three stages recognised first by Deevey (1960) and shown in Fig. 2. There were only two major demographic transitions: (1) from around 46,000 BCE (Before Common Era) to around– 27,000 BCE, characterised by a massive increase in the acceleration of growth, the transition that lasted for around 19,000 years; and (2) from around 425 BCE to around 510 CE, characterised by a massive deceleration, which lasted for around 900 years. The minor disturbance around 1300 CE caused only a small temporary delay in the growth of population without changing substantially the shape of the growth trajectory. Nothing unusual happened around 1950 CE that could be used to mark the beginning of the Anthropocene. There was no intensification of growth. Growth of the world human population started to decelerate irreversibly soon after that year, from the early 1960s CE (US Census Bureau, 2021; World Bank, 2023).

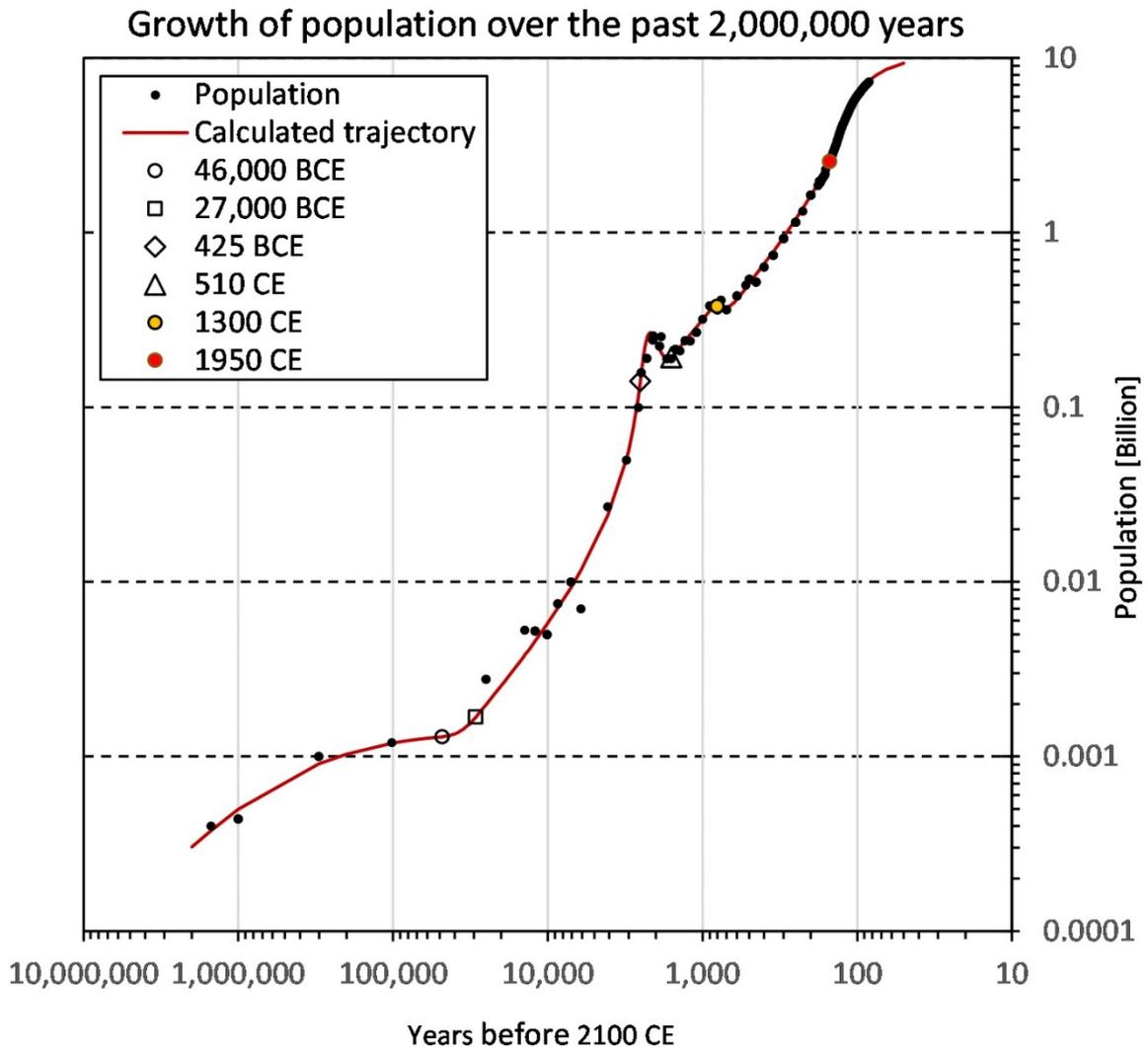

*Figure 2. The largely steady growth of the world human (hominid) population, the primary force of the Anthropocene, suggests a largely gradual development of human skills and a steadily increasing intensity of human impacts on the environment. What is now called the Anthropocene appears to be a part of a long process, a historical event without a manifestation as a new geological epoch because the observed systematic deceleration of anthropogenic growth trajectories strongly suggests a wide-spread decline of anthropogenic forces. Data source: Nielsen (2017b) and references therein*

The current time interval of human existence, described as the Anthropocene, appears to be a natural continuation of a much longer process of human interaction with the environment, a process transcending the Pleistocene and Holocene Epochs. This current stage has been described as "a part of human history", "a new stage in human history", "unique time in human history", "a new and unique time of strong human impacts", and "a new chapter in human existence" (Nielsen, 2021, 2022). The analysis of the long-term growth of human population combined with the analysis of the recent anthropogenic activities and impacts suggests that the Anthropocene is a historical event without a manifestation as a new geological epoch because the observed systematic deceleration of anthropogenic growth trajectories is associated with a systematic decrease in the intensity of anthropogenic forces. This interpretation appears to have been now confirmed by Gibbard, et al. (2021) who proposed that what is now known as the Anthropocene is a geological event rather than a geological epoch.

The placement of the Anthropocene on the time scale is debatable. There is no clear marker for its lower boundary. The beginning of the Anthropocene could be *defined* but there appears to be no clear indication that it can be convincingly and objectively determined. If it is assumed that the beginning of the Anthropocene should be marked by an unusual acceleration of human activities, and that the growth of human population reflects growth of human interaction with the environment, then the only candidate is the acceleration around (35,000 ± 10,000) BCE to a clearly faster trajectory, but this transition lasted for approximately 19,000 years. If the beginning of the Anthropocene is assumed to be marked by the so-called Jinji (human-natural) unconformity (Edgeworth et al., 2015), then the embryonic beginning could be marked by the first use of Oldowan stone tools around 2.5 Ma (million years ago) or even earlier by the Lomekwian stone tool industry (Harmand et al., 2015) around 3.3 Ma.

If it is assumed that its beginning should be placed closer to our time, then maybe some selected data could be used to *define* its lower boundary. For instance, the analysis (Nielsen, 2022) of anthropogenic indicators revealed a strong acceleration of the atmospheric concentration of carbon dioxide in the 1800s CE. Climate change is one of the most critical developments driven by human activities. Maybe this acceleration could be used to mark the beginning of the Anthropocene. Another serious threat is nuclear explosions. Maybe the beginning of the Anthropocene could be marked by the time of the first explosions of nuclear bombs. The precise location of the *assumed* beginning is probably not important. From a long perspective of time, extending over millions of years, singling the 1800s CE or any other recent time for its *proposed* beginning makes no substantial difference.

## Summary

1. **Deceleration**. Deceleration of global change process was expected, and was even described as inevitable (Steffen, et al., 2004). Deceleration was also described as an emergent property of the so-called "Great Acceleration" (Hibbard et al., 2005). It is a part of the same process. It is this deceleration, anticipated and discussed earlier (Steffen et al., 2004; Hibbard et al., 2005), that has now been demonstrated by the analysis of a wide range of anthropogenic growth trajectories both here and in earlier publications (Nielsen, 2021, 2022). Further investigation of this phenomenon, including projections of growth, as outlined earlier (Nielsen, 2021, Suppl.), could help in a better understanding of the Anthropocene and in the identification of areas of the most critical developments.

2. **The "Great Acceleration"**. The so-called "Great Acceleration" is not acceleration (Steffen, 2017, 2018, private communication; Head, at al., 2022). It is *great* because it embraces a wide range of human activities and their strong impacts on the Earth system, but it is not *acceleration*. The term "Great Acceleration" could be perhaps more descriptively and more accurately replaced by the "Great Rapid Increase". However, as pointed out by Head et al. (2022), the term "Great Acceleration" has been used for a long time and it might be hard to replace it by a new term, even if a new term is more accurate. The question is whether the inherent ambiguity of this term can interfere with the correct interpretation of the Anthropocene. If it can, then perhaps it should be changed.

3. **Important difference**. According to Head et al. (2022), the so-called "Great Acceleration" is metaphorical. In contrast, the evidence-based systematic deceleration is literally true. It is a phenomenon confirmed by a wide range of data. The term "Great Rapid Increase", a possible replacement of the term "Great Acceleration", is also literal.

4. **No room for complacency.** The currently experienced systematic deceleration leaves no room for complacency because, in general, anthropogenic trajectories continue to increase. Even after reaching their respective maxima, local or asymptotic, they are likely to continue to be reflected in strong impacts on the Earth System.

5. **1950 CE**. Close inspection of data shows that there was no systematic abrupt acceleration around 1950 CE or around any other recent year. The assumed intensification of growth around 1950 CE is based on impressions but impressions can be misleading, and they are definitely misleading in this case. Exponential and hyperbolic type trajectories, which describe anthropogenic indicators, can be hard to interpret if data are presented using linear scales of reference. A better option is to present them using semilogarithmic scales of reference.

6. **Interpretation of the Anthropocene**. Mathematical analysis of anthropogenic growth trajectories, combined with the analysis of the growth of hominid population over the past 2,000,000 years, suggests that the Anthropocene is not a geological epoch but a historical event. This conclusion is similar to the interpretation proposed now by Gibbard et al. (2022) who suggested that the Anthropocene is not a geological epoch but a geological event. Human footprints of strong environmental impacts can be found in many places, on land (including geological deposits), water (e.g., in critical ocean acidification) and in the atmosphere (notably in the increasing concentration of carbon dioxide). Data suggest that what is now called the Anthropocene is a part of a much longer process of human interaction with nature. The recently experienced strong anthropogenic impacts on the Earth System could have been expected because, on average, growth of human (hominid) population was steadily increasing and even accelerating. Human interaction with the environment was becoming progressively stronger and eventually reached its expected climax. Effects of human impacts on the Earth System might be expected to continue for a long time but there is no clear indication that they can be manifested in a geological transition.

7. **The beginning**. There appears to be no systematic feature in the already analysed data that could be used to *determine* the beginning of the Anthropocene, but maybe its beginning could be *defined*. If it is assumed that the Anthropocene is the most recent phenomenon, then a suitable marker defining its beginning could be perhaps the strong acceleration of the atmospheric concentration of carbon dioxide around the 1800s CE (Nielsen, 2022) or the first use of nuclear bombs, because the continuing increase in the $CO_2$ concentration and the continuing nuclear threat, individually or combined, pose a serious problem to human future and even to the future of our planet. However, the Anthropocene could be also understood as a part of a much longer process transcending the Pleistocene and Holocene epochs, an event without a clearly marked lower boundary. Looking further back in time and using the concept of the Jinji unconformity (Edgeworth et al., 2015), the embryonic beginning of the gradually evolving human (hominid) interaction with the environment could be placed at the time of the first use of stone tools, millions of years ego.

8. **The most important issue**. Interesting as they are, the most important issues are not whether the Anthropocene is a new geological epoch or the beginning of this event, but how to deal with numerous critical environmental problems caused by humans. In particular, one of the most important issues is the ongoing climate change.

9. **Further research**. Further analysis of new data, including extensive investigation of projections of growth, as described earlier (Nielsen, 2021, Suppl.), could assist not only in a better understanding of the Anthropocene and of human impacts on the Earth System but also in deciding which areas of human activities require the most urgent intervention.

# Appendix

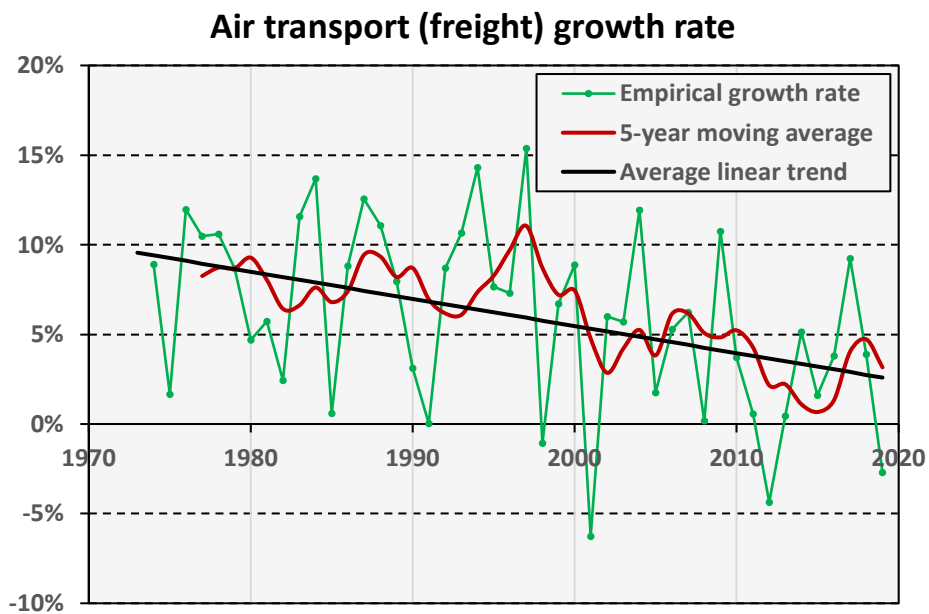

*Figure A1. Growth rate for the global air transport was, on average, decreasing, which means that the growth was gradually slowing down (decelerating).*

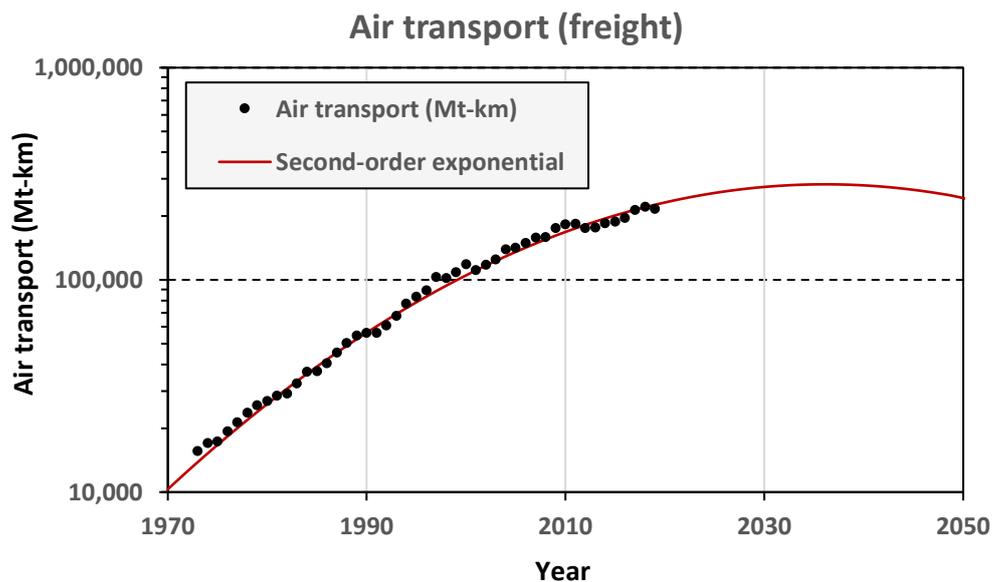

*Figure A2. Growth of the global air transport has been steadily slowing down. The calculated curve is a solution of eqn (1) for the straight line displayed in Fig. A1. In this semilogarithmic display, growth trajectory is bending downwards, which is a signature of a decelerating growth. The projected maximum is around 2036 CE.*

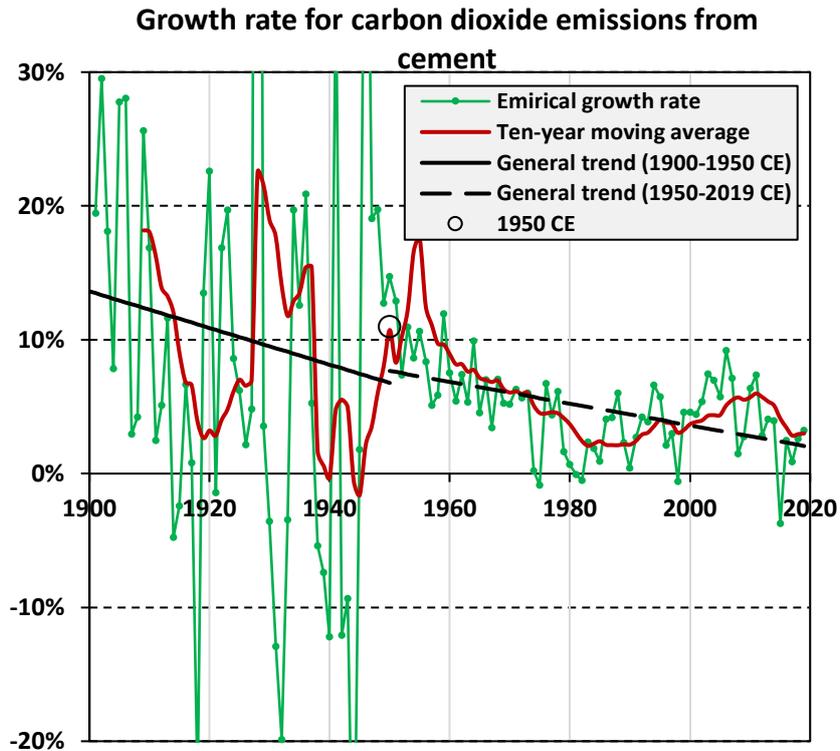

*Figure A3. Growth rate data for the emissions of carbon dioxide from the production of cement identify two distinct regions of growth: a region of large fluctuations before around 1950 CE and a region of significantly smaller fluctuations after 1950 CE. Growth rate was, on average, decreasing in both regions, which means that growth was decelerating. There was no sustained acceleration around 1950 CE.*

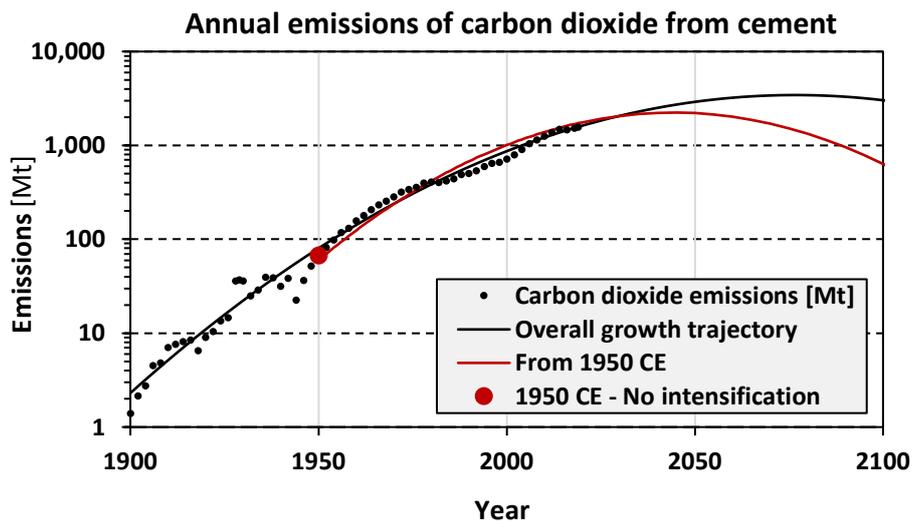

*Figure A4. Growth trajectory is bending downwards and thus demonstrating that the global annual emissions of carbon dioxide from the production of cement was decelerating. Diagrams presented in Figs A3 and A4 illustrate that the general trend of growth is not affected by fluctuations and oscillations in the growth rate but by the general trend of this quantity. The predicted maximum is around 2045 CE for the trajectory generated by growth rate data after 1950 CE, and 2075 CE for the overall trajectory.*

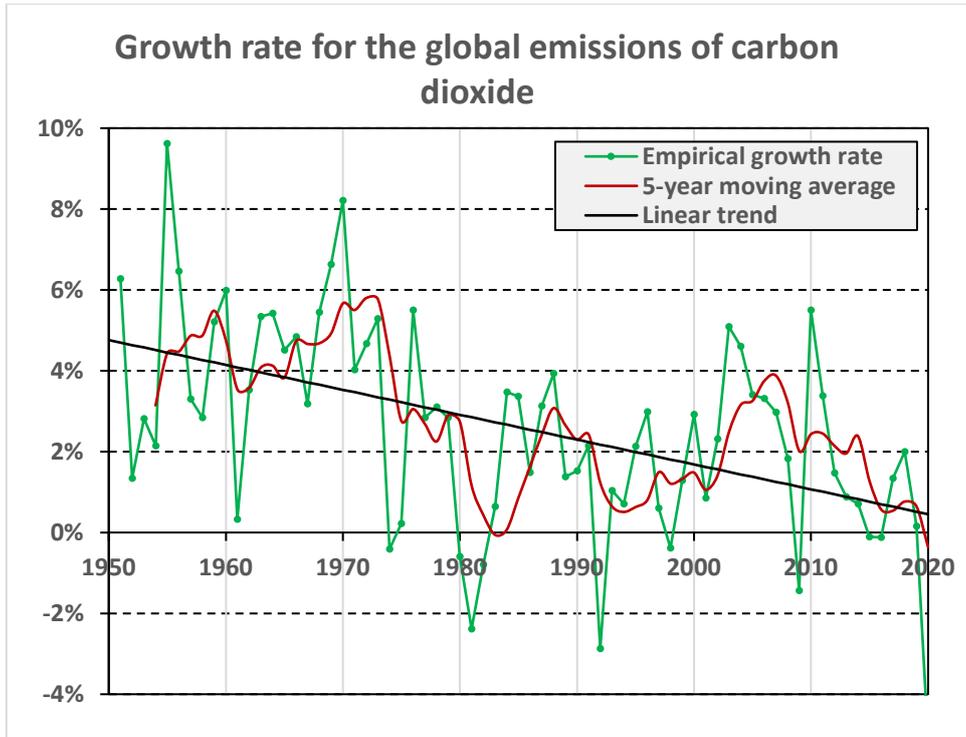

*Figure A5. Growth rate for the annual global emissions of carbon dioxide has been steadily decreasing. Global emissions have been steadily decelerating.*

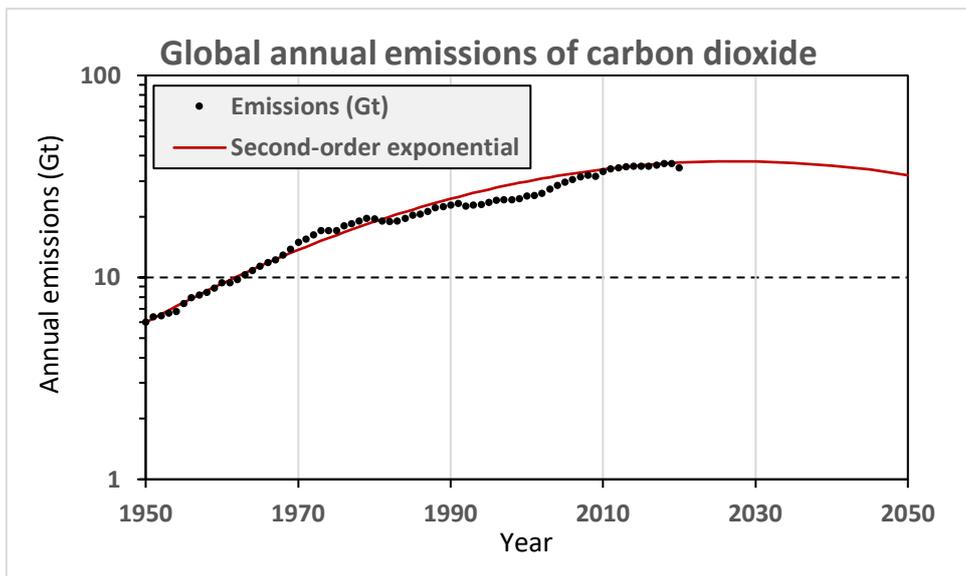

*Figure A6. Global annual emissions of carbon dioxide have been decelerating as indicated here by the trajectory bending downwards in this semilogarithmic display. The second-order exponential trajectory describing data and projecting growth was generated by the straight line shown in Fig. A5 and used in eqn (1).*

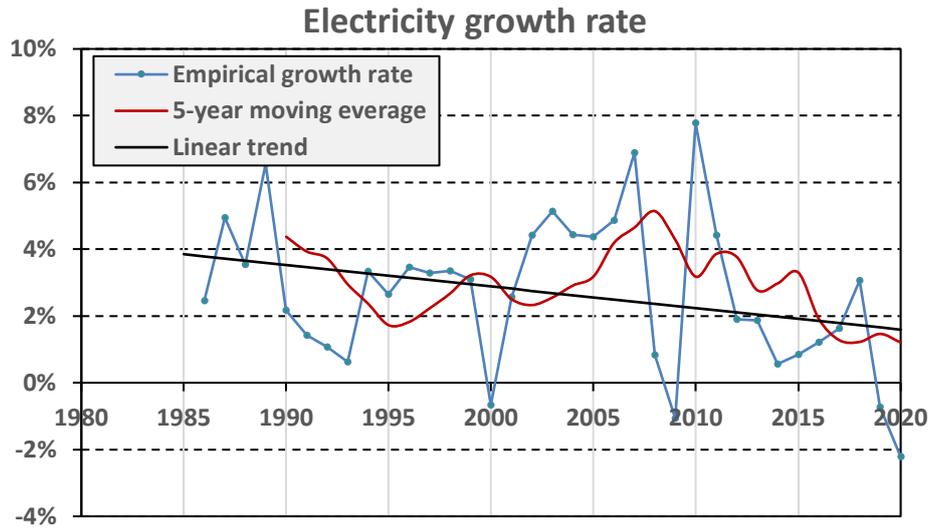

*Figure A7. Growth rate for the annual global consumption of electricity generated by fossil fuels: coal, gas, and oil. On average, growth rate was decreasing. Annual global consumption of electricity generated by fossil fuels has been slowing down.*

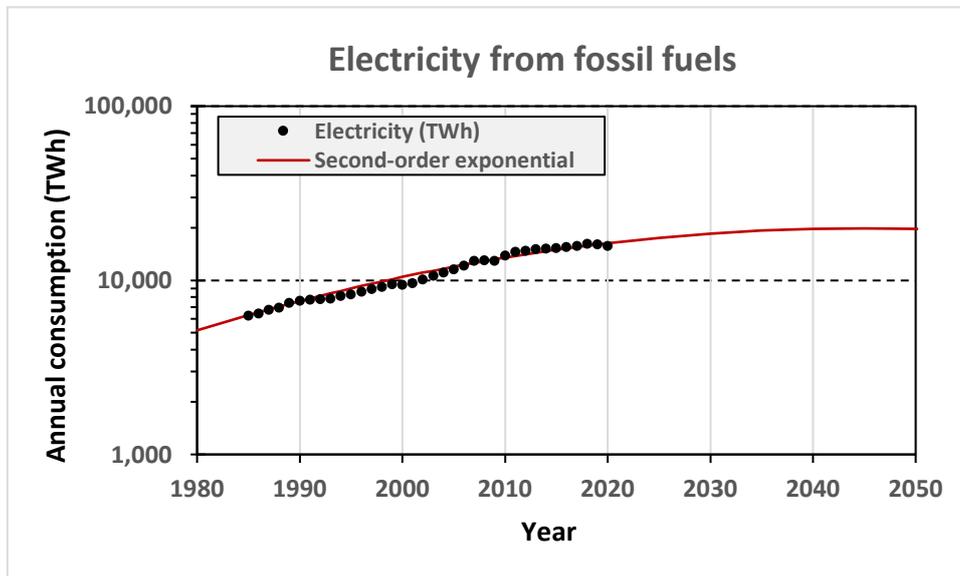

*Figure A8. Global annual consumption of electricity generated by fossil fuels: coal, gas, and oil. The curve describing growth and its projection was generated by the straight line shown in Fig. A7. Consumption of electricity generated by fossil fuels is projected to reach a maximum around 2045 CE but it will remain high for a long time after that year.*

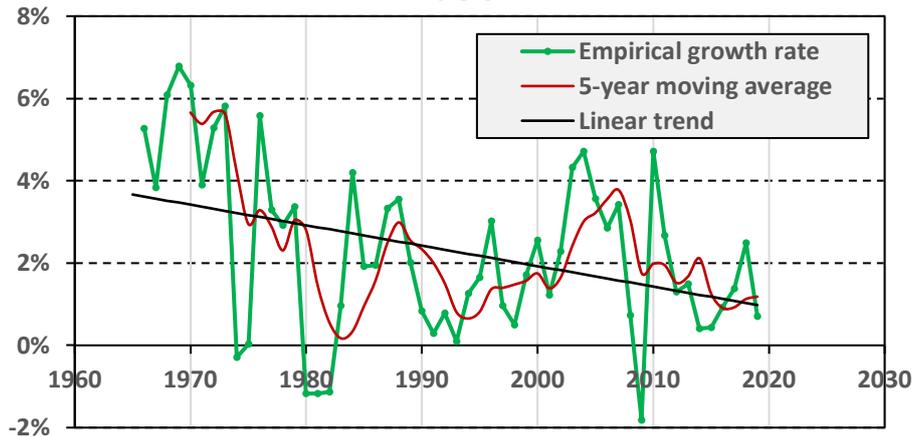

*Figure A9. Growth rate for the annual consumption of primary energy generated by fossil fuels. Consumption of energy from these sources have been gradually slowing down.*

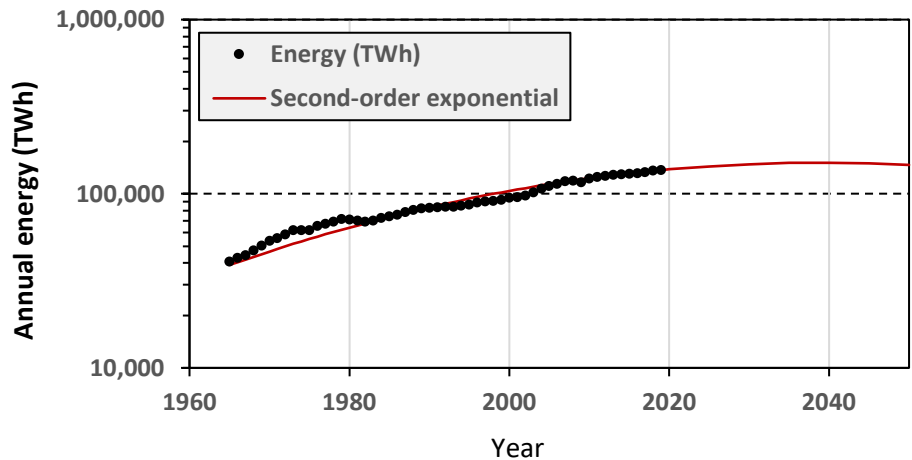

*Figure A10. Global annual consumption of energy generated by fossil fuels. Mathematical description of data and projection of growth was generated by the straight line shown in Fig. A9. If the trend continues, the annual global consumption of energy generated by fossil fuels will reach a maximum around 2040 CE, but it will remain high for a long time after that year.*

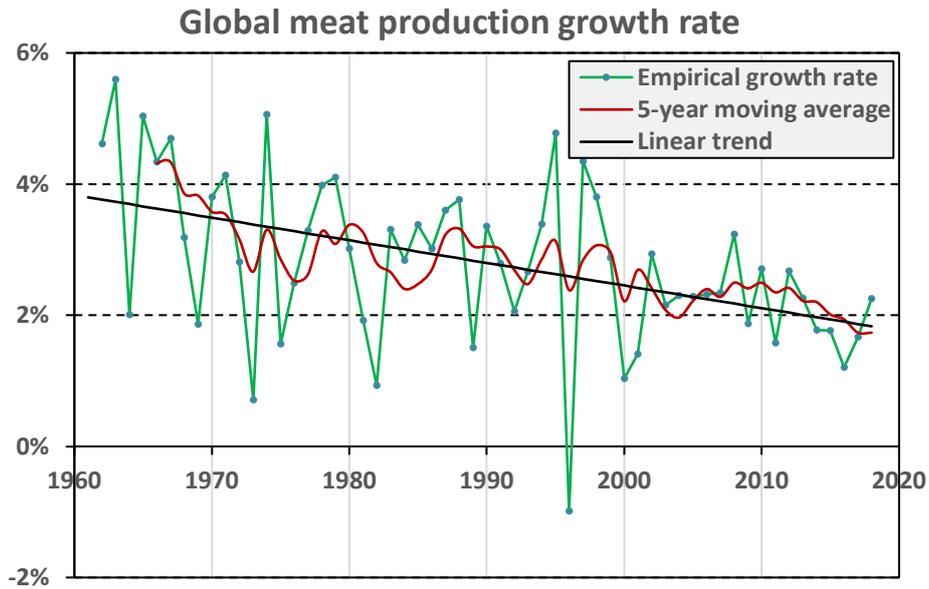

*Figure A11. Growth rate for the global annual production of meat has been steadily decreasing. The annual production has been slowing down.*

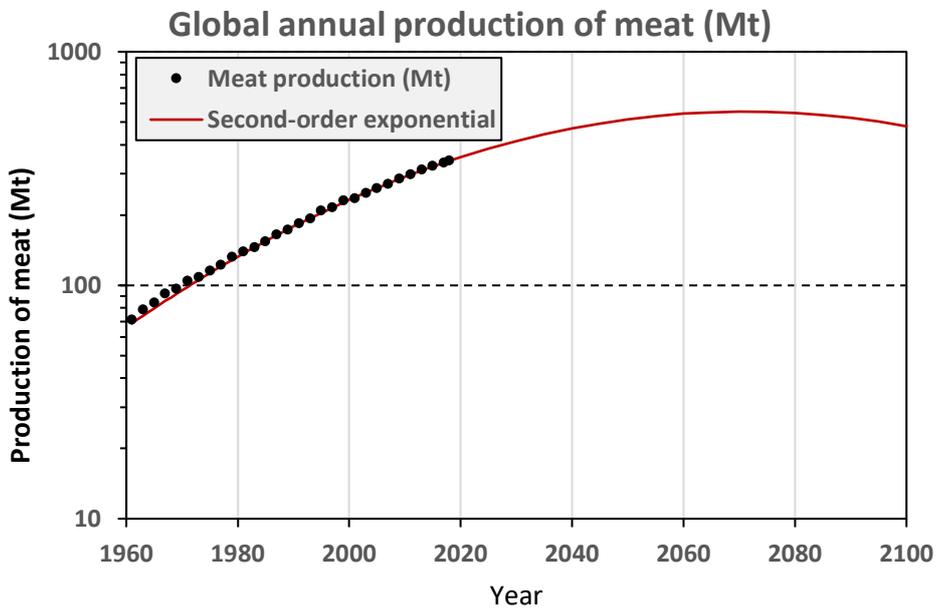

*Figure A12. Global annual production of meat. Mathematical trajectory is a solution of eqn (1) for the straight line shown in Fig. A11. If the trend continues, annual production of meat will reach a maximum in around 2070.*

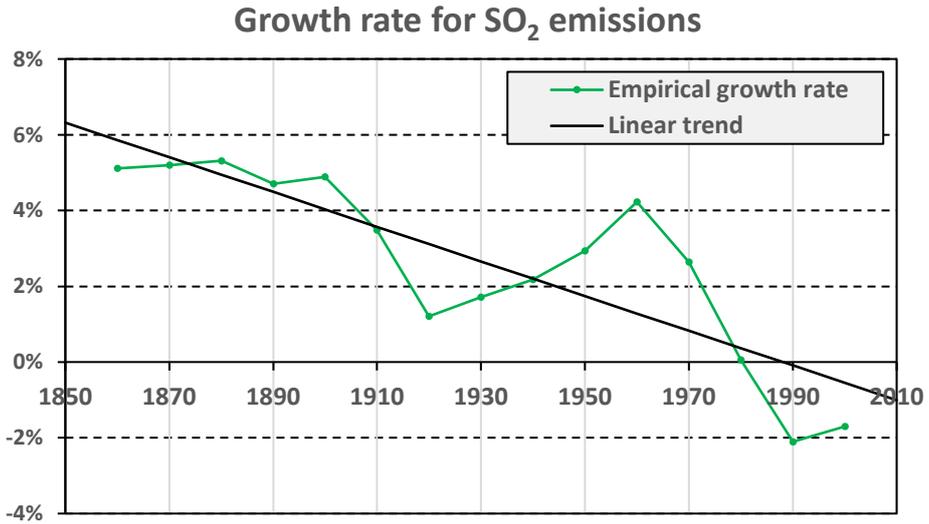

*Figure A14. The average growth rate for the global annual emissions of SO₂. The gradually decreasing growth rate shows that annual emissions of SO₂ have been, on average, gradually slowing down.*

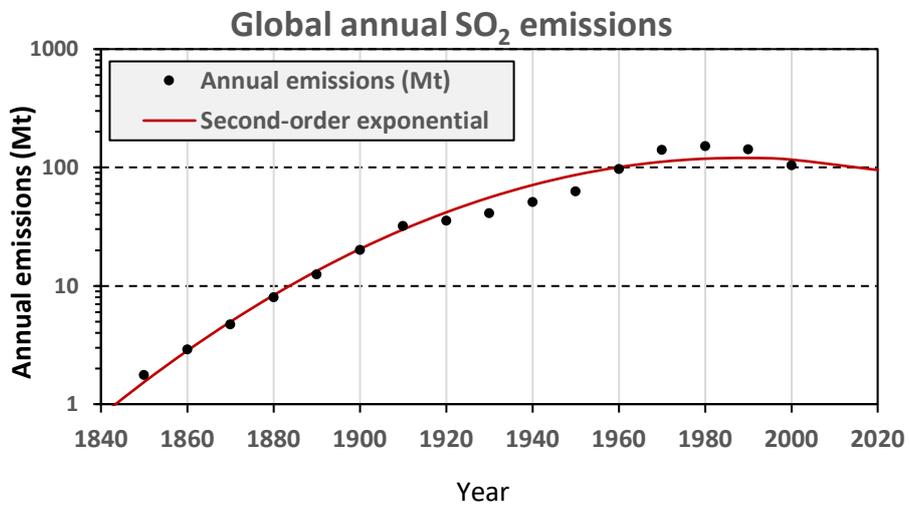

*Figure A14. Global annual emissions of SO₂. Mathematical trajectory is a solution of eqn (1) for the straight line shown in Fig. A13. Annual emissions reached a maximum of 151.5 Mt in 1980 CE and continue to decrease. The trajectory has been decelerating over the entire range of data.*

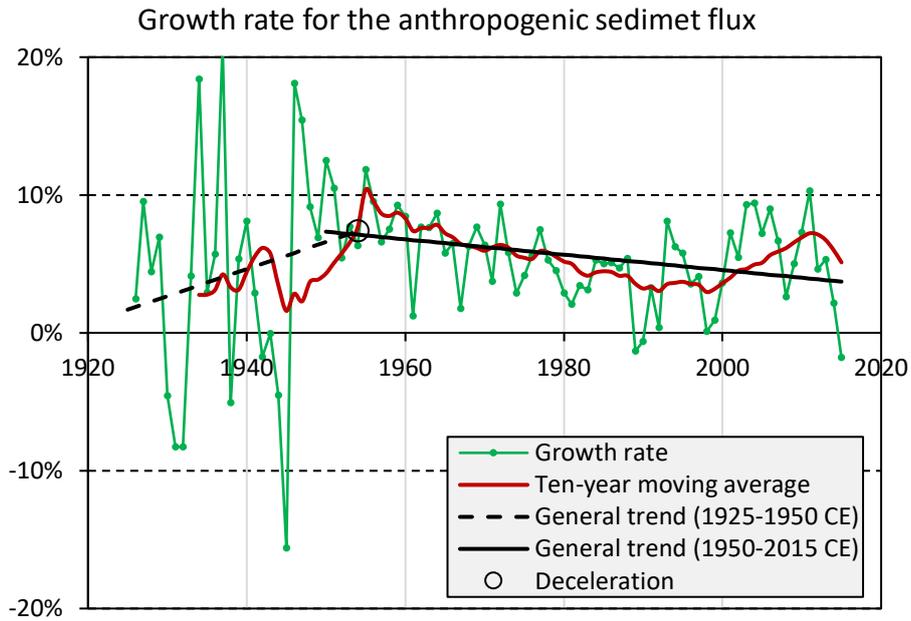

*Figure A15. Growth rate for the total global sediment flux caused by human activities. Growth rate was initially, on average, increasing. Growth was, on average, accelerating. Soon after 1950 CE, growth rate started to decrease. There was no sustained acceleration around 1950 CE but a sustained deceleration. Anthropogenic sediment flux started to decelerate and, on average, continued its declining trend.*

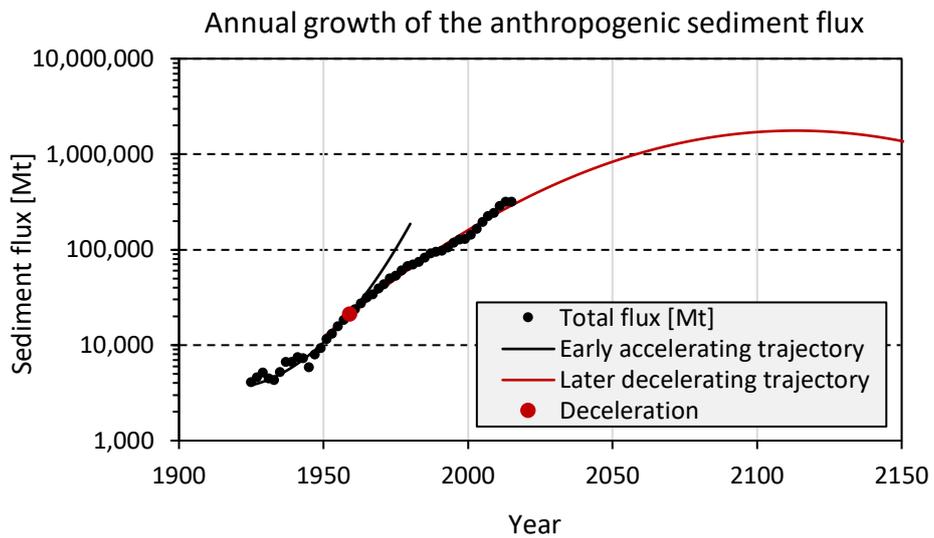

*Figure A16. Global increase of the total annual anthropogenic sediment flux. The presented trajectories were generated by the straight lines shown in Fig. A15. Sediment flux was initially accelerating but soon after 1950 CE it started to decelerate. The predicted maximum is around 2115 CE.*